\documentstyle[12pt,fleqn]{article}

\title{Induced Emission of Gamma Radiation\\
from Isomeric Nuclei\\}

\author{Silviu Olariu and Agata Olariu\\
Institute of Physics and Nuclear Engineering,\\
Department of Fundamental Experimental Physics\\
P.O.Box MG-6, 76900 Magurele, Bucharest, Romania\\
e-mail: olariu@ifin.nipne.ro}

\begin{document}
\date{}
\maketitle

\abstract
We study the possibility to influence the lifetime of nuclear isomeric states 
with the aid of incident fluxes of photons. We assume that a nucleus  
initially in an isomeric state $|i\rangle$ first absorbs an incident 
photon of energy $E_{ni}$ to reach a higher intermediate state $|n\rangle$,
then the state $|n\rangle$ decays to a lower state $|l\rangle$.
In favorable cases the two-step induced emission rates become equal 
to the natural isomeric decay rates for incident power densities of the 
order of $10^{10}$ W cm$^{-2}$.\\

\endabstract

\section{Introduction}

In a recent work it has been shown that the lifetime of isomeric
nuclear states can be influenced by X-ray electron-nuclear double transitions 
(XENDT's), which are processes in which 
a transition effected by an inner atomic electron takes place simultaneously
with a nuclear electromagnetic transition \cite{1}. 
The rate of deexcitation of isomeric nuclei induced by XENDT's was calculated
for the case when the holes in the atomic shells are produced by incident 
ionizing electrons and it was found that the induced nuclear deexcitation rate 
becomes comparable to the natural decay rate for ionizing electron 
fluxes of the order of $10^{14}$ W cm$^{-2}$.  

In this work we shall study the possibility 
to influence the lifetime of nuclear isomeric states with the aid of fluxes of 
gamma-ray photons.
The induced deexcitation of the isomeric nucleus considered 
in this work is a two-step process. We assume that the nucleus  
initially in the isomeric state $|i\rangle$ first absorbs a photon of energy
$E_{ni}$ so that the nucleus reaches the higher intermediate state $|n\rangle$.
The state $|n\rangle$ then decays to a lower state $|l\rangle$ by the emission
of a gamma-ray photon having the energy  $E_{nl}$ or by internal conversion.
In Sec. II we estimate the spectral intensities for which the single-photon
induced emission rates become equal to the isomeric decay rates. These
single-photon intensities turn out to be extremely large.
In Sec. III we introduce the concept of two-step  deexcitation of
isomeric nuclei induced by incident photon fluxes,
and estimate the spectral intensities for which the
two-step induced emission rates become equal to the isomeric decay rates.
In Sec. IV we list the nuclear isomers for which the required incident power
density are within the reach of existing experimental techniques.
In favorable cases 
the two-step induced emission rates become equal to the isomeric decay rates
for incident power densities of the order of $10^{10}$ W cm$^{-2}$.\\

\section{Single-photon induced gamma emission rates}

We consider an isomeric state $|i\rangle$ of energy $E_i$ and assume that a
nucleus in the state $|i\rangle$ decays to a lower state $|l\rangle$ 
of energy $E_l$. A photon of energy $E_{il}$ may be spontaneously 
emitted in this isomeric transition. In the presence of a beam of incident
photons there will also be a certain rate for the {\it induced} emission of
photons of energy $E_{il}$. In this section we estimate this
single-photon induced gamma emission rate. We describe the incident photon flux
by the spectral intensity $N(E)$, defined such that $N(E)dE$ should represent
the number of photons incident per unit surface and time and having the energy
between $E$ and $E+dE$.

The probability per second for the induced emission of a photon 
of energy $E$ is
\begin{equation}
w_i^{(1)}=\int_0^\infty\sigma_i(E)N(E)dE .
\label{61.7}
\end{equation}
The cross section for induced emission $\sigma_i(E)$ is
\begin{equation}
\sigma_i(E)=\frac{\pi c^2\hbar^4}{2E^2}
\frac{\Gamma_{il}(\Gamma_i+\Gamma_l)}{(E-E_{il})^2+\hbar^2(\Gamma_i+
\Gamma_l)^2/4} .
\label{61.8}
\end{equation}
In Eq. (\ref{61.8}) $\Gamma_{il}$ is the partial width for gamma-ray emission
in the transition $|i\rangle\rightarrow |l\rangle$,   $\Gamma_i=\ln 2/t_i$ is 
the natural isomeric decay rate, $t_i$ being the isomeric half-life, and 
$\Gamma_l$ is the total width of the state $|l\rangle$. If $|l\rangle$ is the
ground state, then $\Gamma_l=0$.
The level widths are expressed in this work in units of s$^{-1}$.
If the spectral intensity $N(E)$ is a slowly varying function of $E$ the
single-photon induced emission rate becomes, from Eqs. (\ref{61.7}) and
(\ref{61.8}) 
\begin{equation}
w_i^{(1)}=\frac{\pi^2 c^2\hbar^3}{E_{il}^2}\Gamma_{il} N(E_{il}) .
\label{61.9}
\end{equation}
The half-life $t_i$ of the isomeric state is related to the width $\Gamma_{il}$
by
\begin{equation}
t_i=\frac{\ln 2f_{IT}}{(1+\alpha_{il})\Gamma_{il}} ,
\label{101}
\end{equation}
where $f_{IT}$ is the fraction of isomeric decays by gamma-ray emission 
or electron conversion and
$\alpha_{il}$ is the internal conversion coefficient for the transition
$|i\rangle\rightarrow |l\rangle$.

The single-photon induced-emission rate $w_i^{(1)}$ 
and the  isomeric decay rate $\ln 2/t_i$ become equal for a spectral
intensity 
\begin{equation}
N_1(E_{il})=\frac{(1+\alpha_{il})E_{il}^2}{\pi^2 c^2\hbar^3f_{IT}} .
\label{61.10}
\end{equation}
Numerically we have
\begin{equation}
N_1(E_{il})=3.95 \times 10^{29} \frac{(1+\alpha_{il}) E_{il}^2}{f_{IT}} ,
\label{61.11}
\end{equation}
where $ N_1$ is expressed in 
${\rm photons}\: {\rm cm}^{-2} {\rm s}^{-1} {\rm eV}^{-1}$ 
and $E_{il}$ is expressed in keV. The single-photon 
spectral intensity in Eq. (\ref{61.11}) is extremely large.\\

\section{Two-step deexcitation of isomeric nuclei induced by incident photon
fluxes} 

In this section we analyze the possibility of a two-step deexcitation of
a nuclear isomeric state. The first step consists in the absorption by the
nucleus in the isomeric state $|i\rangle$ of a photon having the energy
$E_{ni}$, the nucleus thus making a transition to a higher
intermediate state $|n\rangle$ of energy $E_n$. 
The state $|n\rangle$ then decays into a lower
state $|l\rangle$ by the emission of a gamma-ray photon having the  energy
$E_{nl}$ or by internal conversion, 
as shown in Fig. 1. In some cases the state $|l\rangle$ may be
situated above the isomeric state $|i\rangle$, and in these cases the
transition $|n\rangle\rightarrow |l\rangle$ is followed by a further gamma-ray
transition to a lower state $|l^\prime\rangle$.

The multipolarity of the single-photon gamma-ray transition
$|i\rangle\rightarrow |l\rangle$ is usually E3, M3, E4 or M4, while the
multipolarity of the transitions $|i\rangle\rightarrow |n\rangle$,
$|n\rangle\rightarrow |l\rangle$ may be lower. Then the rate of the
induced two-step gamma transition 
$|i\rangle\rightarrow |n\rangle\rightarrow |l\rangle$ may be in some cases
higher than the rate of the single-photon induced transition
 $|i\rangle\rightarrow |l\rangle, $
for the same applied spectral intensity. The spin $J_n$ of the nuclear 
intermediate state $|n\rangle$ must be enclosed between the spins
 $J_i, J_l$ of the states $|i\rangle, |l\rangle$. 

For a flux of incident photons of spectral intensity $N(E)$, the rate of the
transition  $|i\rangle\rightarrow|n\rangle$ is
\begin{equation}
w_{in}^{(1)}=\int_0^\infty \frac{2J_n+1}{2J_i+1}
\frac{\pi c^2\hbar^4}{2E^2}
\frac{\Gamma_{ni}\Gamma_n}{(E-E_{ni})^2+\hbar^2\Gamma_n^2/4}N(E)dE ,
\label{62.1}
\end{equation}
where $E_{ni}=E_n-E_i, \Gamma_n $ is the total width of the state $|n\rangle$,
$\Gamma_n\gg \Gamma_i$,
and $\Gamma_{ni}$ is the partial width of the electromagnetic transition
$|n\rangle\rightarrow |i\rangle$. 
The total width $\Gamma_n$ of the intermediate state $|n\rangle$ which decays
into the states $|i\rangle, |l\rangle$ 
and other lower states $|l^\prime \rangle$
is given by
\begin{equation}
\Gamma_n=(1+\alpha_{ni})\Gamma_{ni}+(1+\alpha_{nl})\Gamma_{nl}
+\sum_{l^\prime} (1+\alpha_{nl^\prime})\Gamma_{nl^\prime} ,
\label{62.17}
\end{equation}
where $\alpha_{ni}, \alpha_{nl},  \alpha_{nl^\prime}$ 
are the internal conversion coefficients
for the transitions $|n\rangle\rightarrow |i\rangle,
|n\rangle\rightarrow |l\rangle, |n\rangle\rightarrow |l^\prime \rangle.$  
The rate $w_{il}^{(2)}$ for the two-step
transition $|i\rangle\rightarrow |n\rangle\rightarrow |l\rangle$ is then
\begin{equation}
w_{il}^{(2)}=w_{in}^{(1)}(1+\alpha_{nl})\Gamma_{nl}/\Gamma_n ,
\label{62.2}
\end{equation}
where $\Gamma_{nl}$ is the partial width of the electromagnetic transition
$|n\rangle\rightarrow |l\rangle$. If $N(E)$ 
is a slowly-varying function of energy, the two-step transition rate is
\begin{equation}
w_{il}^{(2)}=\frac{2J_n+1}{2J_i+1}\frac{\pi^2 c^2\hbar^3}{E_{ni}^2}
\frac{(1+\alpha_{nl})\Gamma_{ni}\Gamma_{nl}}{\Gamma_n}N(E_{ni}).
\label{62.3}
\end{equation}
The two-step transition rate is thus proportional to the effective width 
\begin{equation}
\Gamma_{eff}=(1+\alpha_{nl})\Gamma_{ni}\Gamma_{nl}/\Gamma_n.
\label{103}
\end{equation} 
The two-step induced rate $w_{il}^{(2)}$ becomes equal to the isomeric decay
rate $\ln 2/t_i$ for a spectral intensity
\begin{equation}
N_2(E_{ni})=\frac{2J_i+1}{2J_n+1}\frac{E_{ni}^2}{\pi^2 c^2\hbar^3}
\frac{1}{F}\frac{t_n}{t_i} ,
\label{62.4}
\end{equation}
where
\begin{equation}
F=(1+\alpha_{nl})\Gamma_{ni}\Gamma_{nl}/\Gamma_n^2 .
\label{62.4b}
\end{equation}
In Eq. (\ref{62.4b}) the width $\Gamma_n$
is related to the half-life $t_n$ of the state $|n\rangle$ by $\Gamma_n=\ln
2/t_n$. We have $1/F\geq 4$, but usually $1/F\gg 4$.

The ratio $N_2(E_{ni})/N_1(E_{nl})$ is
\begin{equation}
N_2(E_{ni})/N_1(E_{il})=\frac{2J_i+1}{2J_n+1}
\frac{f_{IT}}{(1+\alpha_{il})}
\frac{E_{ni}^2}{E_{il}^2}
\frac{1}{F}\frac{t_n}{t_i} .
\label{102}
\end{equation}
Although usually $1/F \gg 4$, we have however $t_n/t_i\ll 1$, and 
$N_2/N_1\ll 1$. Thus, the two-step approach to the problem of induced gamma
emission requires much lower incident power densities than the single-photon  
approach.

The number of
photons per unit surface and time in the incident beam is of the order of
$N_2(E_{ni}) E_{ni}$, and the power density, measured in W cm$^{-2}$, 
is of the order of 
\begin{equation}
P_2=N_2(E_{ni}) E_{ni}^2. 
\label{100}
\end{equation}
We have numerically
\begin{equation}
N_2(E_{ni})=3.95\times 10^{29}
\frac{2J_i+1}{2J_n+1}\frac{\Gamma_n^2}{(1+\alpha_{nl})
\Gamma_{ni}\Gamma_{nl}} \frac{t_n}{t_i}E_{ni}^2 ,
\label{62.6}
\end{equation}
\begin{equation}
P_2=6.33\times 10^{16}
\frac{2J_i+1}{2J_n+1}\frac{\Gamma_n^2}{(1+\alpha_{nl})
\Gamma_{ni}\Gamma_{nl}} \frac{t_n}{t_i}E_{ni}^4 ,
\label{104}
\end{equation}
where $N_2$ is expressed in 
${\rm photons}\: {\rm cm}^{-2} {\rm s}^{-1} {\rm eV}^{-1}$, 
$P_2$ in W cm$^{-2}$ and $E_{ni}$ in keV.
For $\Gamma_n^2/\Gamma_{ni}\Gamma_{nl}$=4, 
$\alpha_{nl}=0$, $t_n/t_i=10^{-13}, E_{ni}=30$ keV, 
$J_i=J_n$ we have $N_2=1.42\times 10^{20}$  
${\rm photons}\: {\rm cm}^{-2} {\rm s}^{-1} {\rm eV}^{-1}$, and
$N_2 E_{ni}=4.27\times 10^{24} {\rm photons}\: {\rm cm}^{-2} {\rm s}^{-1} $, 
which for $E_{ni}$=30 keV corresponds to $N_2 E_{ni}^2=2.05\times 10^{10}$ 
W cm$^{-2}$, which
is not exceedingly high. However, as mentioned previously, the 
quantity $\Gamma_n^2/(1+\alpha_{nl})\Gamma_{ni}\Gamma_{nl}$ 
has values which in reality 
are much larger than the lower limit of 4. \\

\section{Case study}

We have determined the spectral intensity $N_2$ and 
the power level $P_2$ at which the induced emission rate
becomes equal to the isomeric decay rate for isomeric nuclei
having a half-life $t_i>$10 min and for which the cascade originating on 
the state $|n\rangle$ contains a gamma-ray transition of energy 
$E_\gamma >2 E_{ni}$. The latter relation represents 
a condition of upconversion of the energy of the incident photons.  
Some of the isomeric nuclei having these properties are listed in Table I.
In the analysis of the nuclear properties we have used the Table of Isotopes
\cite{2}, and the internal conversion coefficients are taken from the BNL
data base \cite{3} and from refs. \cite{4} and \cite{5}. 
If $t_n, t_i$ and the relative gamma-ray intensities 
$R_{ni}, R_{nl}, R_{nl^\prime}$ 
of the downward transitions from the state $|n\rangle$ are all known, we have
calculated $N_2$ and $P_2$ from Eqs. (\ref{62.4}), (\ref{62.4b})
and (\ref{100}), using for
$F$ the value
\begin{equation}
F_R=\frac{(1+\alpha_{nl})R_{ni}R_{nl}}
{\left[(1+\alpha_{ni})R_{ni}+(1+\alpha_{nl})R_{nl}
+\sum_{l^\prime} (1+\alpha_{nl^\prime})R_{nl^\prime}\right]^2} \:.
\end{equation}
Otherwise we have
calculated $N_2$ and $P_2$ by using the
Weisskopf estimates for the radiative widths of the transitions and
the half-lives $t_n, t_i$ appearing in Eqs. (\ref{62.4}), (\ref{62.4b}) 
and (\ref{100}). The 
error of the values of $N_2$ and $P_2$ calculated with the aid of the 
Weisskopf estimates is in general of about two orders of magnitude.

We see from Table I that the lowest values of $P_2$ are of the order of
$10^{10}$ W cm$^{-2}$, while in most cases the gamma-ray power density 
$P_2$ is of the order of $10^{14}$ W cm$^{-2}$. In the case of the 6.8 h 
isomeric nucleus $^{93}$Mo, a 4.8 keV photon generates a
cascade in which an 1477 keV photon is emitted. In the case of 
the 3.5 h isomeric nucleus $^{202}$Pb, a 38.6 keV photon generates a cascade
in which a 960 keV photon is emitted. 
In the case of the 67.2 min isomeric nucleus $^{204}$Pb, a 78.56 keV
photon produces the emission of a 990 keV photon. 
In the case of the 141 y isomeric
nucleus $^{242}$Am, a 4.3 keV photon produces the emission of a 52.9 keV
photon. 
In the case of the 2.9 y isomeric
nucleus $^{102}$Rh, a 13.7 keV photon produces the emission of a 112.5 keV
photon. 
In the case of the 13.7 h
isomeric nucleus $^{69}$Zn, a 92.7 keV photon produces the emission of a 531
keV photon. In the case of the 6.01 h isomeric nucleus 
$^{99}$Tc, a 38.4 keV photon produces the emission of a 181.0 keV photon. 
In the case of the 31 y isomeric nucleus $^{178}$Hf, an 126.1 keV
photon generates a cascade in which a 574 keV photon is emitted.

The analysis has been restricted in this work to cases of upconversion, 
for which the cascade originating on the state $|n\rangle$ contains a 
gamma-ray photon of energy $E_\gamma >2 E_{ni}$. Moreover, the analysis has
been restricted to isomeric nuclei for which there is a tabulated intermediate 
state $|n\rangle$ of known energy and known spin and parity. 
In the proximity of isomeric states there are also tabulated levels of known 
energy but unknown or uncertain spin or parity. The assignement of spins 
and parities and the discovery of new states may lead in time to the finding 
of isomeric nuclei with favorable intermediate states for which 
the power level $P_2$ should be lower than the present estimate of $10^{10}$ 
W cm$^{-2}$.

The power level $P_2$ required for the observation of induced gamma emission
is large because the effective width $\Gamma_{eff}$ appearing in Eq.
(\ref{103}) is a very small fraction of the total width of the 
spectrum of pumping radiation. The use of small-area isomeric samples and of
pulsed incident radiation may reduce the total energy of the incident pulse to
practical values. An alternative to the broad-band pumping is the use of
M\"{o}ssbauer sources of gamma radiation. Considering an isomeric nucleus of
proton number $Z$ and mass number $A$, the nuclei $(Z-1,A)$ or $(Z+1,A)$ may in
certain cases populate the intermediate state $|n\rangle$ of the nucleus 
$(A,Z)$ by $\beta^-$ or respectively $\beta^+,$ EC disintegrations. In this
way, M\"{o}ssbauer sources containing the nuclei $(Z-1,A)$ or $(Z+1,A)$ may 
produce photon fluxes of energy $E_{ni}$ in a narrow interval of width
$\hbar \Gamma_n$ around $E_{ni}$. An example of this kind is 
$_{\:\:82}^{202}$Pb, whose 42 ns intermediate state at 2208.4 keV is populated 
by the decay of the nucleus $^{202}_{\:\:83}$Bi, which has a half-life of 
1.72 h. On the other hand, while M\"{o}ssbauer sources can produce large 
spectral intensities, they cannot produce pulses of radiation. The two-step 
transition rates obtained from the M\"{o}ssbauer scattering by isomeric nuclei 
depends much on the practical details of the experiment. \\

\section{Conclusions}

The power levels required by 
the photon approach to the problem of induced gamma emission studied in this
work are comparable to the power levels required by the 
XENDT approach described in \cite{1}. The choice of isomeric nuclei was
restricted in this work by the upconversion condition $E_\gamma > 2E_{ni}$,
and it was restricted to intermediate states of known energy and known spin and
parity. For the cases considered in this work, the required level of incident 
power for which the two-step emission rate becomes equal to the natural
emission rate of isomeric nuclei is of the order of $10^{10}$ W cm$^{-2}$. \\

Acknowledgments\\

This work has been supported by a research grant from the Romanian Academy 
of Sciences. \\

\newpage

Fig. 1.  Paths for the deexcitation of a nuclear isomeric state.
The two possibilities are the spontaneous deexcitation of the 
isomeric state $|i\rangle$ to a lower state $|l\rangle$, and the two-step 
deexcitation, when the nucleus first makes a transition from the 
isomeric state $|i\rangle$ to a higher nuclear 
state $|n\rangle$ by the absorption of a photon of energy $E_{ni}$,
then it emits a photon to reach the lower nuclear state $|l\rangle$.

\newpage

Table I. Spectral intensity $N_2$ and 
power level $P_2$ for which the induced emission rate becomes equal to
the natural emission rate from an isomeric nucleus. The isomeric level has the
energy $E_i$ and the half-life $t_i$, the incident photons have the energy
$E_{ni}$, $E_{nl}$ is the energy of the gamma-ray photon emitted from the 
state 
$|n\rangle$ and $E_\gamma$ is the largest photon energy in the cascade from the
intermediate state $|n\rangle$. An R in the method column means that 
$N_2$ and  $P_2$ have been calculated  from
the relative gamma-ray intensities and the measured $t_n$ and $t_i$, and 
a W in the method column means that $N_2$ and $P_2$ have been obtained from
the Weisskopf estimates for the gamma-ray widths and half-lives.

\newpage

\begin{tabular}{|c|r|c|r|r|c|c|c|r|}
\hline
Nucleus&$E_i$,&$t_i$, &$E_{ni}$,&$E_{nl}$,&$N_2,$                        
&$P_2$,                       &Method&$E_\gamma/E_{ni}$\\
       &keV   &       &keV      &keV      
&${\rm ph./(cm^{2}\: s \:eV)}$&${\rm W/cm^2}$  &      &       \\   
\hline
$^{186}$Re	&149.0	&2.0$\times 10^{5}$ y	
&37.0	& 12.1& 5.1$\times 10^{19}$ &1.1$\times 10^{10}$     &W&2.0\\
$^{152}$Eu      &45.6	&9.27 h          	
&32.6	& 13.0& $5.8\times 10^{21}$ &1.0$\times 10^{12}$     &R&2.0\\
$^{52}$Mn	&377.7	&21.1 m         	
&353.7	&731.5& $1.1\times 10^{21}$ &2.1$\times 10^{13}$     &R&2.1\\
$^{152}$Eu	&45.6	&9.27 h           	
&19.7	&65.3 & 6.3$\times 10^{23}$ &3.9$\times 10^{13}$     &R&3.3\\
$^{178}$Hf	&2446.1	&31 y            	
&126.1	&140.3& 4.4$\times 10^{22}$ &1.1$\times 10^{14}$     &R&4.6\\
$^{96}$Tc       &34.3	&51.5 m          	
&11.1	& 45.3& 6.8$\times 10^{24}$ &1.3$\times 10^{14}$     &W&4.1\\
$^{44}$Sc	&271.1	&58.6 h          	
&78.7	&349.8& 2.4$\times 10^{23}$ &2.4$\times 10^{14}$     &W&4.4\\
$^{202}$Pb	&2169.8	&3.53 h          	
&38.6	&168.1& 1.3$\times 10^{24}$ &3.1$\times 10^{14}$     &W&17.0\\
$^{99}$Tc       &142.7    &6.01 h                 
&38.4     &181.1& 1.6$\times 10^{24}$ &3.7$\times 10^{14}$     &R&4.7\\
$^{178}$Hf      &2446.1   &31 y                   
&126.1    &437.0& 1.5$\times 10^{23}$ &3.9$\times 10^{14}$     &R&4.2\\
$^{99}$Tc       &142.7    &6.01 h                 
&38.4     & 40.6& 2.2$\times 10^{24}$ &5.2$\times 10^{14}$     &R&3.7\\
$^{119}$Te      &261.0	&4.70 d          	
&99.4	& 40.0& $6.5\times 10^{23}$ &1.0$\times 10^{15}$     &W&3.2\\
$^{71}$Zn	&157.7	&3.96 h           	
&128.6	&286.3& 4.7$\times 10^{23}$ &1.3$\times 10^{15}$     &W&2.2\\
$^{201}$Bi      &846.3	&59.1 m         	
&240.1	&195.9& 4.6$\times 10^{23}$ &4.3$\times 10^{15}$     &R&3.7\\
$^{242}$Am	&48.6	&141 y          	
&4.3	& 52.9& 2.1$\times 10^{27}$ &6.1$\times 10^{15}$     &W&12.4\\
$^{93}$Mo	&2424.9	&6.85 h          	
&4.8	&267.9& 2.8$\times 10^{27}$ &1.0$\times 10^{16}$     &W&307.7\\
$^{186}$Re      &149.0	&2.0$\times 10^{5}$ y   
&37.0	& 86.6&$9.5\times 10^{25}$  &2.1$\times 10^{16}$      &W&2.3\\
$^{69}$Zn	&438.6	&13.7 h         	
&92.7	&531.2& 2.2$\times 10^{25}$ &3.0$\times 10^{16}$     &W&5.7\\
$^{84}$Rb	&463.6	&20.2 m          	
&151.4	&615.0& 9.1$\times 10^{24}$ &3.4$\times 10^{16}$     &W&4.1\\
$^{204}$Pb	&2185.8	&67.2 m         	
&78.6	&990.3& 5.0$\times 10^{25}$ &5.0$\times 10^{16}$     &R&12.6\\
$^{115}$In	&336.2	&4.48 h           	
&260.9	&597.1& 5.9$\times 10^{24}$ &6.4$\times 10^{16}$     &W&2.3\\
$^{191}$Os	&74.4	&13.1 h         	
&57.6	&131.9& 1.6$\times 10^{26}$ &8.5$\times 10^{16}$     &W&2.3\\
$^{96}$Tc	&34.3	&51.5 m          	
&14.9	& 49.2& 2.5$\times 10^{27}$ &9.0$\times 10^{16}$     &W&3.3\\
$^{102}$Rh	&140.8	&2.9 y           	
&13.7	&112.5& 9.4$\times 10^{27}$ &2.8$\times 10^{17}$     &W&8.2\\
\hline
\end{tabular}

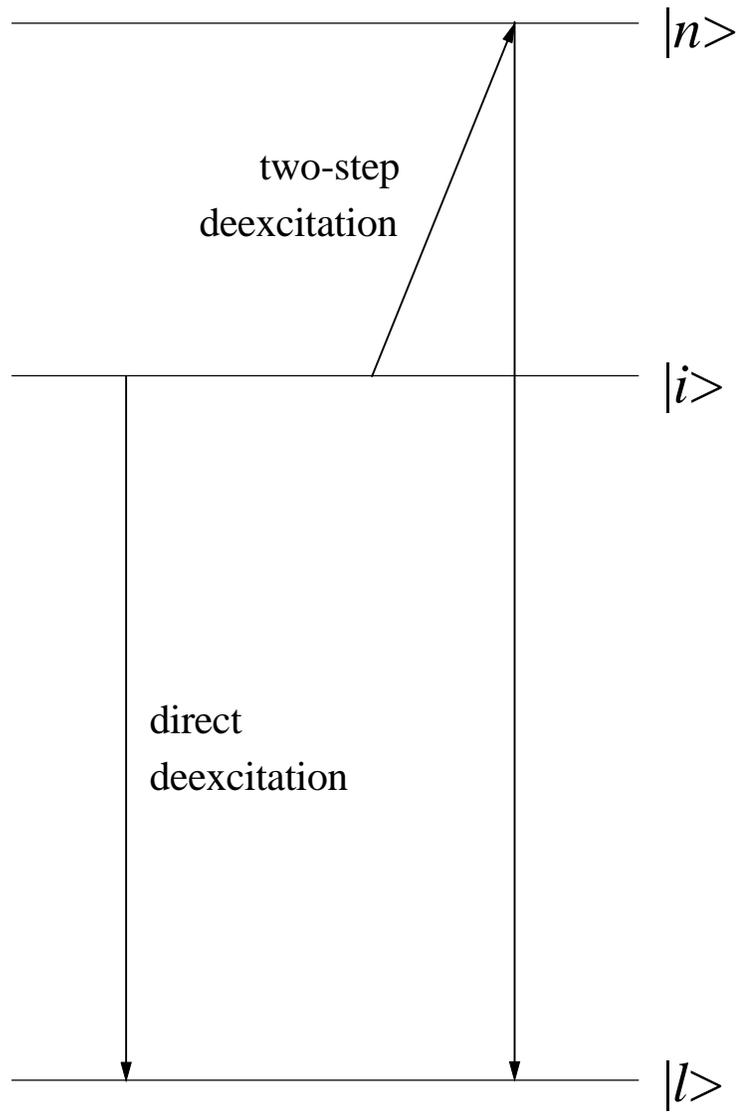
\end{document}